\documentclass{ws-procs9x6}
\usepackage{graphicx}
\usepackage[dvips]{color} 
\usepackage{epsfig}
\usepackage{amssymb,amsmath}
\usepackage{longtable}
\usepackage[breaklinks, colorlinks, citecolor=blue]{hyperref}
\usepackage{breakurl}
\usepackage{multirow}
\usepackage{url}
\usepackage{txfonts}
\DeclareGraphicsExtensions{.eps,.ps,.jpg,.png}
\def\deg{$^\circ$}

\def\setsymbol#1#2{\expandafter\def\csname #1\endcsname{#2}}
\def\getsymbol#1{\csname #1\endcsname}

\def\Planck{{\it Planck\/}}

\newbox\tablebox    \newdimen\tablewidth
\def\leaderfil{\leaders\hbox to 5pt{\hss.\hss}\hfil}
\def\tablenote#1 #2\par{\begingroup \parindent=0.8em
    \abovedisplayshortskip=0pt\belowdisplayshortskip=0pt
    \noindent
    $$\hss\vbox{\hsize\tablewidth \hangindent=\parindent \hangafter=1 \noindent
    \hbox to \parindent{\sup{\rm #1}\hss}\strut#2\strut\par}\hss$$
    \endgroup}

\def\L2{\ifmmode L_2\else $L_2$\fi}

\def\DeltaT{\ifmmode \Delta T\else $\Delta T$\fi}
\def\deltat{\ifmmode \Delta t\else $\Delta t$\fi}
\def\fknee{\ifmmode f_{\rm knee}\else $f_{\rm knee}$\fi}
\def\Fmax{\ifmmode F_{\rm max}\else $F_{\rm max}$\fi}
\def\solar{\ifmmode{\rm M}_{\mathord\odot}\else${\rm M}_{\mathord\odot}$\fi}

\def\inv{\ifmmode^{-1}\else$^{-1}$\fi}
\def\mo{\ifmmode^{-1}\else$^{-1}$\fi}
\def\sup#1{\ifmmode ^{\rm #1}\else $^{\rm #1}$\fi}
\def\expo#1{\ifmmode \times 10^{#1}\else $\times 10^{#1}$\fi}
\def\,{\thinspace}
\def\lsim{\mathrel{\raise .4ex\hbox{\rlap{$<$}\lower 1.2ex\hbox{$\sim$}}}}
\def\gsim{\mathrel{\raise .4ex\hbox{\rlap{$>$}\lower 1.2ex\hbox{$\sim$}}}}

\def\simprop{\mathrel{\raise .4ex\hbox{\rlap{$\propto$}\lower 1.2ex\hbox{$\sim$}}}}
\def\deg{\ifmmode^\circ\else$^\circ$\fi}
\def\pdeg{\ifmmode $\setbox0=\hbox{$^{\circ}$}\rlap{\hskip.11\wd0 .}$^{\circ}
          \else \setbox0=\hbox{$^{\circ}$}\rlap{\hskip.11\wd0 .}$^{\circ}$\fi}
\def\arcs{\ifmmode {^{\scriptstyle\prime\prime}}
          \else $^{\scriptstyle\prime\prime}$\fi}
\def\arcm{\ifmmode {^{\scriptstyle\prime}}
          \else $^{\scriptstyle\prime}$\fi}
\newdimen\sa  \newdimen\sb
\def\parcs{\sa=.07em \sb=.03em
     \ifmmode \hbox{\rlap{.}}^{\scriptstyle\prime\kern -\sb\prime}\hbox{\kern -\sa}
     \else \rlap{.}$^{\scriptstyle\prime\kern -\sb\prime}$\kern -\sa\fi}
\def\parcm{\sa=.08em \sb=.03em
     \ifmmode \hbox{\rlap{.}\kern\sa}^{\scriptstyle\prime}\hbox{\kern-\sb}
     \else \rlap{.}\kern\sa$^{\scriptstyle\prime}$\kern-\sb\fi}
\def\ra[#1 #2 #3.#4]{#1\sup{h}#2\sup{m}#3\sup{s}\llap.#4}
\def\dec[#1 #2 #3.#4]{#1\deg#2\arcm#3\arcs\llap.#4}
\def\deco[#1 #2 #3]{#1\deg#2\arcm#3\arcs}
\def\rra[#1 #2]{#1\sup{h}#2\sup{m}}

\def\dots{\relax\ifmmode \ldots\else $\ldots$\fi}
\def\WHzsr{\ifmmode $W\,Hz\mo\,sr\mo$\else W\,Hz\mo\,sr\mo\fi}
\def\mHz{\ifmmode $\,mHz$\else \,mHz\fi}
\def\GHz{\ifmmode $\,GHz$\else \,GHz\fi}
\def\mKs{\ifmmode $\,mK\,s$^{1/2}\else \,mK\,s$^{1/2}$\fi}
\def\muKs{\ifmmode \,\mu$K\,s$^{1/2}\else \,$\mu$K\,s$^{1/2}$\fi}
\def\muKRJs{\ifmmode \,\mu$K$_{\rm RJ}$\,s$^{1/2}\else \,$\mu$K$_{\rm RJ}$\,s$^{1/2}$\fi}
\def\muKCMBs{\ifmmode \,\mu$K$_{\rm CMB}$\,s$^{1/2}\else \,$\mu$K$_{\rm CMB}$\,s$^{1/2}$\fi}
\def\muKHz{\ifmmode \,\mu$K\,Hz$^{-1/2}\else \,$\mu$K\,Hz$^{-1/2}$\fi}
\def\MJysr{\ifmmode \,$MJy\,sr\mo$\else \,MJy\,sr\mo\fi}
\def\MJysrmK{\ifmmode \,$MJy\,sr\mo$\,mK$_{\rm CMB}\mo\else \,MJy\,sr\mo\,mK$_{\rm CMB}\mo$\fi}
\def\microns{\ifmmode \,\mu$m$\else \,$\mu$m\fi}

\def\muK{\ifmmode \,\mu$K$\else \,$\mu$\hbox{K}\fi}
\def\microK{\ifmmode \,\mu$K$\else \,$\mu$\hbox{K}\fi}
\def\muW{\ifmmode \,\mu$W$\else \,$\mu$\hbox{W}\fi}
\def\kms{\ifmmode $\,km\,s$^{-1}\else \,km\,s$^{-1}$\fi}
\def\kmsMpc{\ifmmode $\,\kms\,Mpc\mo$\else \,\kms\,Mpc\mo\fi}

\setsymbol{LFI:center:frequency:70GHz:units}{70.3\,GHz}
\setsymbol{LFI:center:frequency:44GHz:units}{44.1\,GHz}
\setsymbol{LFI:center:frequency:30GHz:units}{28.5\,GHz}

\setsymbol{LFI:center:frequency:70GHz}{70.3}
\setsymbol{LFI:center:frequency:44GHz}{44.1}
\setsymbol{LFI:center:frequency:30GHz}{28.5}

\setsymbol{LFI:center:frequency:LFI18:Rad:M:units}{71.7\GHz}
\setsymbol{LFI:center:frequency:LFI19:Rad:M:units}{67.5\GHz}
\setsymbol{LFI:center:frequency:LFI20:Rad:M:units}{69.2\GHz}
\setsymbol{LFI:center:frequency:LFI21:Rad:M:units}{70.4\GHz}
\setsymbol{LFI:center:frequency:LFI22:Rad:M:units}{71.5\GHz}
\setsymbol{LFI:center:frequency:LFI23:Rad:M:units}{70.8\GHz}
\setsymbol{LFI:center:frequency:LFI24:Rad:M:units}{44.4\GHz}
\setsymbol{LFI:center:frequency:LFI25:Rad:M:units}{44.0\GHz}
\setsymbol{LFI:center:frequency:LFI26:Rad:M:units}{43.9\GHz}
\setsymbol{LFI:center:frequency:LFI27:Rad:M:units}{28.3\GHz}
\setsymbol{LFI:center:frequency:LFI28:Rad:M:units}{28.8\GHz}
\setsymbol{LFI:center:frequency:LFI18:Rad:S:units}{70.1\GHz}
\setsymbol{LFI:center:frequency:LFI19:Rad:S:units}{69.6\GHz}
\setsymbol{LFI:center:frequency:LFI20:Rad:S:units}{69.5\GHz}
\setsymbol{LFI:center:frequency:LFI21:Rad:S:units}{69.5\GHz}
\setsymbol{LFI:center:frequency:LFI22:Rad:S:units}{72.8\GHz}
\setsymbol{LFI:center:frequency:LFI23:Rad:S:units}{71.3\GHz}
\setsymbol{LFI:center:frequency:LFI24:Rad:S:units}{44.1\GHz}
\setsymbol{LFI:center:frequency:LFI25:Rad:S:units}{44.1\GHz}
\setsymbol{LFI:center:frequency:LFI26:Rad:S:units}{44.1\GHz}
\setsymbol{LFI:center:frequency:LFI27:Rad:S:units}{28.5\GHz}
\setsymbol{LFI:center:frequency:LFI28:Rad:S:units}{28.2\GHz}

\setsymbol{LFI:center:frequency:LFI18:Rad:M}{71.7}
\setsymbol{LFI:center:frequency:LFI19:Rad:M}{67.5}
\setsymbol{LFI:center:frequency:LFI20:Rad:M}{69.2}
\setsymbol{LFI:center:frequency:LFI21:Rad:M}{70.4}
\setsymbol{LFI:center:frequency:LFI22:Rad:M}{71.5}
\setsymbol{LFI:center:frequency:LFI23:Rad:M}{70.8}
\setsymbol{LFI:center:frequency:LFI24:Rad:M}{44.4}
\setsymbol{LFI:center:frequency:LFI25:Rad:M}{44.0}
\setsymbol{LFI:center:frequency:LFI26:Rad:M}{43.9}
\setsymbol{LFI:center:frequency:LFI27:Rad:M}{28.3}
\setsymbol{LFI:center:frequency:LFI28:Rad:M}{28.8}
\setsymbol{LFI:center:frequency:LFI18:Rad:S}{70.1}
\setsymbol{LFI:center:frequency:LFI19:Rad:S}{69.6}
\setsymbol{LFI:center:frequency:LFI20:Rad:S}{69.5}
\setsymbol{LFI:center:frequency:LFI21:Rad:S}{69.5}
\setsymbol{LFI:center:frequency:LFI22:Rad:S}{72.8}
\setsymbol{LFI:center:frequency:LFI23:Rad:S}{71.3}
\setsymbol{LFI:center:frequency:LFI24:Rad:S}{44.1}
\setsymbol{LFI:center:frequency:LFI25:Rad:S}{44.1}
\setsymbol{LFI:center:frequency:LFI26:Rad:S}{44.1}
\setsymbol{LFI:center:frequency:LFI27:Rad:S}{28.5}
\setsymbol{LFI:center:frequency:LFI28:Rad:S}{28.2}

\setsymbol{LFI:white:noise:sensitivity:70GHz:units}{134.7\muKs}
\setsymbol{LFI:white:noise:sensitivity:44GHz:units}{164.7\muKs}
\setsymbol{LFI:white:noise:sensitivity:30GHz:units}{143.4\muKs}

\setsymbol{LFI:white:noise:sensitivity:70GHz}{134.7}
\setsymbol{LFI:white:noise:sensitivity:44GHz}{164.7}
\setsymbol{LFI:white:noise:sensitivity:30GHz}{143.4}

\setsymbol{LFI:white:noise:sensitivity:LFI18:Rad:M:units}{512.0\muKs}
\setsymbol{LFI:white:noise:sensitivity:LFI19:Rad:M:units}{581.4\muKs}
\setsymbol{LFI:white:noise:sensitivity:LFI20:Rad:M:units}{590.8\muKs}
\setsymbol{LFI:white:noise:sensitivity:LFI21:Rad:M:units}{455.2\muKs}
\setsymbol{LFI:white:noise:sensitivity:LFI22:Rad:M:units}{492.0\muKs}
\setsymbol{LFI:white:noise:sensitivity:LFI23:Rad:M:units}{507.7\muKs}
\setsymbol{LFI:white:noise:sensitivity:LFI24:Rad:M:units}{462.2\muKs}
\setsymbol{LFI:white:noise:sensitivity:LFI25:Rad:M:units}{413.6\muKs}
\setsymbol{LFI:white:noise:sensitivity:LFI26:Rad:M:units}{478.6\muKs}
\setsymbol{LFI:white:noise:sensitivity:LFI27:Rad:M:units}{277.7\muKs}
\setsymbol{LFI:white:noise:sensitivity:LFI28:Rad:M:units}{312.3\muKs}
\setsymbol{LFI:white:noise:sensitivity:LFI18:Rad:S:units}{465.7\muKs}
\setsymbol{LFI:white:noise:sensitivity:LFI19:Rad:S:units}{555.6\muKs}
\setsymbol{LFI:white:noise:sensitivity:LFI20:Rad:S:units}{623.2\muKs}
\setsymbol{LFI:white:noise:sensitivity:LFI21:Rad:S:units}{564.1\muKs}
\setsymbol{LFI:white:noise:sensitivity:LFI22:Rad:S:units}{534.4\muKs}
\setsymbol{LFI:white:noise:sensitivity:LFI23:Rad:S:units}{542.4\muKs}
\setsymbol{LFI:white:noise:sensitivity:LFI24:Rad:S:units}{399.2\muKs}
\setsymbol{LFI:white:noise:sensitivity:LFI25:Rad:S:units}{392.6\muKs}
\setsymbol{LFI:white:noise:sensitivity:LFI26:Rad:S:units}{418.6\muKs}
\setsymbol{LFI:white:noise:sensitivity:LFI27:Rad:S:units}{302.9\muKs}
\setsymbol{LFI:white:noise:sensitivity:LFI28:Rad:S:units}{285.3\muKs}

\setsymbol{LFI:white:noise:sensitivity:LFI18:Rad:M}{512.0}
\setsymbol{LFI:white:noise:sensitivity:LFI19:Rad:M}{581.4}
\setsymbol{LFI:white:noise:sensitivity:LFI20:Rad:M}{590.8}
\setsymbol{LFI:white:noise:sensitivity:LFI21:Rad:M}{455.2}
\setsymbol{LFI:white:noise:sensitivity:LFI22:Rad:M}{492.0}
\setsymbol{LFI:white:noise:sensitivity:LFI23:Rad:M}{507.7}
\setsymbol{LFI:white:noise:sensitivity:LFI24:Rad:M}{462.2}
\setsymbol{LFI:white:noise:sensitivity:LFI25:Rad:M}{413.6}
\setsymbol{LFI:white:noise:sensitivity:LFI26:Rad:M}{478.6}
\setsymbol{LFI:white:noise:sensitivity:LFI27:Rad:M}{277.7}
\setsymbol{LFI:white:noise:sensitivity:LFI28:Rad:M}{312.3}
\setsymbol{LFI:white:noise:sensitivity:LFI18:Rad:S}{465.7}
\setsymbol{LFI:white:noise:sensitivity:LFI19:Rad:S}{555.6}
\setsymbol{LFI:white:noise:sensitivity:LFI20:Rad:S}{623.2}
\setsymbol{LFI:white:noise:sensitivity:LFI21:Rad:S}{564.1}
\setsymbol{LFI:white:noise:sensitivity:LFI22:Rad:S}{534.4}
\setsymbol{LFI:white:noise:sensitivity:LFI23:Rad:S}{542.4}
\setsymbol{LFI:white:noise:sensitivity:LFI24:Rad:S}{399.2}
\setsymbol{LFI:white:noise:sensitivity:LFI25:Rad:S}{392.6}
\setsymbol{LFI:white:noise:sensitivity:LFI26:Rad:S}{418.6}
\setsymbol{LFI:white:noise:sensitivity:LFI27:Rad:S}{302.9}
\setsymbol{LFI:white:noise:sensitivity:LFI28:Rad:S}{285.3}

\setsymbol{LFI:knee:frequency:70GHz:units}{29.5\mHz}
\setsymbol{LFI:knee:frequency:44GHz:units}{56.2\mHz}
\setsymbol{LFI:knee:frequency:30GHz:units}{113.7\mHz}

\setsymbol{LFI:knee:frequency:70GHz}{29.5}
\setsymbol{LFI:knee:frequency:44GHz}{56.2}
\setsymbol{LFI:knee:frequency:30GHz}{113.7}

\setsymbol{LFI:knee:frequency:LFI18:Rad:M:units}{16.3\mHz}
\setsymbol{LFI:knee:frequency:LFI19:Rad:M:units}{15.1\mHz}
\setsymbol{LFI:knee:frequency:LFI20:Rad:M:units}{18.7\mHz}
\setsymbol{LFI:knee:frequency:LFI21:Rad:M:units}{37.2\mHz}
\setsymbol{LFI:knee:frequency:LFI22:Rad:M:units}{12.7\mHz}
\setsymbol{LFI:knee:frequency:LFI23:Rad:M:units}{34.6\mHz}
\setsymbol{LFI:knee:frequency:LFI24:Rad:M:units}{46.2\mHz}
\setsymbol{LFI:knee:frequency:LFI25:Rad:M:units}{24.9\mHz}
\setsymbol{LFI:knee:frequency:LFI26:Rad:M:units}{67.6\mHz}
\setsymbol{LFI:knee:frequency:LFI27:Rad:M:units}{187.4\mHz}
\setsymbol{LFI:knee:frequency:LFI28:Rad:M:units}{122.2\mHz}
\setsymbol{LFI:knee:frequency:LFI18:Rad:S:units}{17.7\mHz}
\setsymbol{LFI:knee:frequency:LFI19:Rad:S:units}{22.0\mHz}
\setsymbol{LFI:knee:frequency:LFI20:Rad:S:units}{8.7\mHz}
\setsymbol{LFI:knee:frequency:LFI21:Rad:S:units}{25.9\mHz}
\setsymbol{LFI:knee:frequency:LFI22:Rad:S:units}{15.8\mHz}
\setsymbol{LFI:knee:frequency:LFI23:Rad:S:units}{129.8\mHz}
\setsymbol{LFI:knee:frequency:LFI24:Rad:S:units}{100.9\mHz}
\setsymbol{LFI:knee:frequency:LFI25:Rad:S:units}{38.9\mHz}
\setsymbol{LFI:knee:frequency:LFI26:Rad:S:units}{58.9\mHz}
\setsymbol{LFI:knee:frequency:LFI27:Rad:S:units}{104.4\mHz}
\setsymbol{LFI:knee:frequency:LFI28:Rad:S:units}{40.7\mHz}

\setsymbol{LFI:knee:frequency:LFI18:Rad:M}{16.3}
\setsymbol{LFI:knee:frequency:LFI19:Rad:M}{15.1}
\setsymbol{LFI:knee:frequency:LFI20:Rad:M}{18.7}
\setsymbol{LFI:knee:frequency:LFI21:Rad:M}{37.2}
\setsymbol{LFI:knee:frequency:LFI22:Rad:M}{12.7}
\setsymbol{LFI:knee:frequency:LFI23:Rad:M}{34.6}
\setsymbol{LFI:knee:frequency:LFI24:Rad:M}{46.2}
\setsymbol{LFI:knee:frequency:LFI25:Rad:M}{24.9}
\setsymbol{LFI:knee:frequency:LFI26:Rad:M}{67.6}
\setsymbol{LFI:knee:frequency:LFI27:Rad:M}{187.4}
\setsymbol{LFI:knee:frequency:LFI28:Rad:M}{122.2}
\setsymbol{LFI:knee:frequency:LFI18:Rad:S}{17.7}
\setsymbol{LFI:knee:frequency:LFI19:Rad:S}{22.0}
\setsymbol{LFI:knee:frequency:LFI20:Rad:S}{8.7}
\setsymbol{LFI:knee:frequency:LFI21:Rad:S}{25.9}
\setsymbol{LFI:knee:frequency:LFI22:Rad:S}{15.8}
\setsymbol{LFI:knee:frequency:LFI23:Rad:S}{129.8}
\setsymbol{LFI:knee:frequency:LFI24:Rad:S}{100.9}
\setsymbol{LFI:knee:frequency:LFI25:Rad:S}{38.9}
\setsymbol{LFI:knee:frequency:LFI26:Rad:S}{58.9}
\setsymbol{LFI:knee:frequency:LFI27:Rad:S}{104.4}
\setsymbol{LFI:knee:frequency:LFI28:Rad:S}{40.7}

\setsymbol{LFI:slope:70GHz:units}{$-1.03$\mHz}
\setsymbol{LFI:slope:44GHz:units}{$-0.89$\mHz}
\setsymbol{LFI:slope:30GHz:units}{$-0.87$\mHz}

\setsymbol{LFI:slope:70GHz}{$-1.03$}
\setsymbol{LFI:slope:44GHz}{$-0.89$}
\setsymbol{LFI:slope:30GHz}{$-0.87$}

\setsymbol{LFI:slope:LFI18:Rad:M:units}{$-1.04$\mHz}
\setsymbol{LFI:slope:LFI19:Rad:M:units}{$-1.09$\mHz}
\setsymbol{LFI:slope:LFI20:Rad:M:units}{$-0.69$\mHz}
\setsymbol{LFI:slope:LFI21:Rad:M:units}{$-1.56$\mHz}
\setsymbol{LFI:slope:LFI22:Rad:M:units}{$-1.01$\mHz}
\setsymbol{LFI:slope:LFI23:Rad:M:units}{$-0.96$\mHz}
\setsymbol{LFI:slope:LFI24:Rad:M:units}{$-0.83$\mHz}
\setsymbol{LFI:slope:LFI25:Rad:M:units}{$-0.91$\mHz}
\setsymbol{LFI:slope:LFI26:Rad:M:units}{$-0.95$\mHz}
\setsymbol{LFI:slope:LFI27:Rad:M:units}{$-0.87$\mHz}
\setsymbol{LFI:slope:LFI28:Rad:M:units}{$-0.88$\mHz}
\setsymbol{LFI:slope:LFI18:Rad:S:units}{$-1.15$\mHz}
\setsymbol{LFI:slope:LFI19:Rad:S:units}{$-1.00$\mHz}
\setsymbol{LFI:slope:LFI20:Rad:S:units}{$-0.95$\mHz}
\setsymbol{LFI:slope:LFI21:Rad:S:units}{$-0.92$\mHz}
\setsymbol{LFI:slope:LFI22:Rad:S:units}{$-1.01$\mHz}
\setsymbol{LFI:slope:LFI23:Rad:S:units}{$-0.95$\mHz}
\setsymbol{LFI:slope:LFI24:Rad:S:units}{$-0.73$\mHz}
\setsymbol{LFI:slope:LFI25:Rad:S:units}{$-1.16$\mHz}
\setsymbol{LFI:slope:LFI26:Rad:S:units}{$-0.79$\mHz}
\setsymbol{LFI:slope:LFI27:Rad:S:units}{$-0.82$\mHz}
\setsymbol{LFI:slope:LFI28:Rad:S:units}{$-0.91$\mHz}

\setsymbol{LFI:slope:LFI18:Rad:M}{$-1.04$}
\setsymbol{LFI:slope:LFI19:Rad:M}{$-1.09$}
\setsymbol{LFI:slope:LFI20:Rad:M}{$-0.69$}
\setsymbol{LFI:slope:LFI21:Rad:M}{$-1.56$}
\setsymbol{LFI:slope:LFI22:Rad:M}{$-1.01$}
\setsymbol{LFI:slope:LFI23:Rad:M}{$-0.96$}
\setsymbol{LFI:slope:LFI24:Rad:M}{$-0.83$}
\setsymbol{LFI:slope:LFI25:Rad:M}{$-0.91$}
\setsymbol{LFI:slope:LFI26:Rad:M}{$-0.95$}
\setsymbol{LFI:slope:LFI27:Rad:M}{$-0.87$}
\setsymbol{LFI:slope:LFI28:Rad:M}{$-0.88$}
\setsymbol{LFI:slope:LFI18:Rad:S}{$-1.15$}
\setsymbol{LFI:slope:LFI19:Rad:S}{$-1.00$}
\setsymbol{LFI:slope:LFI20:Rad:S}{$-0.95$}
\setsymbol{LFI:slope:LFI21:Rad:S}{$-0.92$}
\setsymbol{LFI:slope:LFI22:Rad:S}{$-1.01$}
\setsymbol{LFI:slope:LFI23:Rad:S}{$-0.95$}
\setsymbol{LFI:slope:LFI24:Rad:S}{$-0.73$}
\setsymbol{LFI:slope:LFI25:Rad:S}{$-1.16$}
\setsymbol{LFI:slope:LFI26:Rad:S}{$-0.79$}
\setsymbol{LFI:slope:LFI27:Rad:S}{$-0.82$}
\setsymbol{LFI:slope:LFI28:Rad:S}{$-0.91$}

\setsymbol{LFI:FWHM:70GHz:units}{13\parcm01}
\setsymbol{LFI:FWHM:44GHz:units}{27\parcm92}
\setsymbol{LFI:FWHM:30GHz:units}{32\parcm65}

\setsymbol{LFI:FWHM:70GHz}{13.01}
\setsymbol{LFI:FWHM:44GHz}{27.92}
\setsymbol{LFI:FWHM:30GHz}{32.65}

\setsymbol{LFI:FWHM:LFI18:units}{13\parcm39}
\setsymbol{LFI:FWHM:LFI19:units}{13\parcm01}
\setsymbol{LFI:FWHM:LFI20:units}{12\parcm75}
\setsymbol{LFI:FWHM:LFI21:units}{12\parcm74}
\setsymbol{LFI:FWHM:LFI22:units}{12\parcm87}
\setsymbol{LFI:FWHM:LFI23:units}{13\parcm27}
\setsymbol{LFI:FWHM:LFI24:units}{22\parcm98}
\setsymbol{LFI:FWHM:LFI25:units}{30\parcm46}
\setsymbol{LFI:FWHM:LFI26:units}{30\parcm31}
\setsymbol{LFI:FWHM:LFI27:units}{32\parcm65}
\setsymbol{LFI:FWHM:LFI28:units}{32\parcm66}

\setsymbol{LFI:FWHM:LFI18}{13.39}
\setsymbol{LFI:FWHM:LFI19}{13.01}
\setsymbol{LFI:FWHM:LFI20}{12.75}
\setsymbol{LFI:FWHM:LFI21}{12.74}
\setsymbol{LFI:FWHM:LFI22}{12.87}
\setsymbol{LFI:FWHM:LFI23}{13.27}
\setsymbol{LFI:FWHM:LFI24}{22.98}
\setsymbol{LFI:FWHM:LFI25}{30.46}
\setsymbol{LFI:FWHM:LFI26}{30.31}
\setsymbol{LFI:FWHM:LFI27}{32.65}
\setsymbol{LFI:FWHM:LFI28}{32.66}

\setsymbol{LFI:FWHM:uncertainty:LFI18:units}{0.170\arcm}
\setsymbol{LFI:FWHM:uncertainty:LFI19:units}{0.174\arcm}
\setsymbol{LFI:FWHM:uncertainty:LFI20:units}{0.170\arcm}
\setsymbol{LFI:FWHM:uncertainty:LFI21:units}{0.156\arcm}
\setsymbol{LFI:FWHM:uncertainty:LFI22:units}{0.164\arcm}
\setsymbol{LFI:FWHM:uncertainty:LFI23:units}{0.171\arcm}
\setsymbol{LFI:FWHM:uncertainty:LFI24:units}{0.652\arcm}
\setsymbol{LFI:FWHM:uncertainty:LFI25:units}{1.075\arcm}
\setsymbol{LFI:FWHM:uncertainty:LFI26:units}{1.131\arcm}
\setsymbol{LFI:FWHM:uncertainty:LFI27:units}{1.266\arcm}
\setsymbol{LFI:FWHM:uncertainty:LFI28:units}{1.287\arcm}

\setsymbol{LFI:FWHM:uncertainty:LFI18}{0.170}
\setsymbol{LFI:FWHM:uncertainty:LFI19}{0.174}
\setsymbol{LFI:FWHM:uncertainty:LFI20}{0.170}
\setsymbol{LFI:FWHM:uncertainty:LFI21}{0.156}
\setsymbol{LFI:FWHM:uncertainty:LFI22}{0.164}
\setsymbol{LFI:FWHM:uncertainty:LFI23}{0.171}
\setsymbol{LFI:FWHM:uncertainty:LFI24}{0.652}
\setsymbol{LFI:FWHM:uncertainty:LFI25}{1.075}
\setsymbol{LFI:FWHM:uncertainty:LFI26}{1.131}
\setsymbol{LFI:FWHM:uncertainty:LFI27}{1.266}
\setsymbol{LFI:FWHM:uncertainty:LFI28}{1.287}

\setsymbol{HFI:center:frequency:100GHz:units}{100\,GHz}
\setsymbol{HFI:center:frequency:143GHz:units}{143\,GHz}
\setsymbol{HFI:center:frequency:217GHz:units}{217\,GHz}
\setsymbol{HFI:center:frequency:353GHz:units}{353\,GHz}
\setsymbol{HFI:center:frequency:545GHz:units}{545\,GHz}
\setsymbol{HFI:center:frequency:857GHz:units}{857\,GHz}

\setsymbol{HFI:center:frequency:100GHz}{100}
\setsymbol{HFI:center:frequency:143GHz}{143}
\setsymbol{HFI:center:frequency:217GHz}{217}
\setsymbol{HFI:center:frequency:353GHz}{353}
\setsymbol{HFI:center:frequency:545GHz}{545}
\setsymbol{HFI:center:frequency:857GHz}{857}

\setsymbol{HFI:Ndetectors:100GHz}{8}
\setsymbol{HFI:Ndetectors:143GHz}{11}
\setsymbol{HFI:Ndetectors:217GHz}{12}
\setsymbol{HFI:Ndetectors:353GHz}{12}
\setsymbol{HFI:Ndetectors:545GHz}{3}
\setsymbol{HFI:Ndetectors:857GHz}{4}

\setsymbol{HFI:FWHM:Maps:100GHz:units}{9\parcm88}
\setsymbol{HFI:FWHM:Maps:143GHz:units}{7\parcm18}
\setsymbol{HFI:FWHM:Maps:217GHz:units}{4\parcm87}
\setsymbol{HFI:FWHM:Maps:353GHz:units}{4\parcm65}
\setsymbol{HFI:FWHM:Maps:545GHz:units}{4\parcm72}
\setsymbol{HFI:FWHM:Maps:857GHz:units}{4\parcm39}
\setsymbol{HFI:FWHM:Maps:100GHz}{9.88}
\setsymbol{HFI:FWHM:Maps:143GHz}{7.18}
\setsymbol{HFI:FWHM:Maps:217GHz}{4.87}
\setsymbol{HFI:FWHM:Maps:353GHz}{4.65}
\setsymbol{HFI:FWHM:Maps:545GHz}{4.72}
\setsymbol{HFI:FWHM:Maps:857GHz}{4.39}

\setsymbol{HFI:beam:ellipticity:Maps:100GHz}{1.15}
\setsymbol{HFI:beam:ellipticity:Maps:143GHz}{1.01}
\setsymbol{HFI:beam:ellipticity:Maps:217GHz}{1.06}
\setsymbol{HFI:beam:ellipticity:Maps:353GHz}{1.05}
\setsymbol{HFI:beam:ellipticity:Maps:545GHz}{1.14}
\setsymbol{HFI:beam:ellipticity:Maps:857GHz}{1.19}

\setsymbol{HFI:FWHM:Mars:100GHz:units}{9\parcm37}
\setsymbol{HFI:FWHM:Mars:143GHz:units}{7\parcm04}
\setsymbol{HFI:FWHM:Mars:217GHz:units}{4\parcm68}
\setsymbol{HFI:FWHM:Mars:353GHz:units}{4\parcm43}
\setsymbol{HFI:FWHM:Mars:545GHz:units}{3\parcm80}
\setsymbol{HFI:FWHM:Mars:857GHz:units}{3\parcm67}

\setsymbol{HFI:FWHM:Mars:100GHz}{9.37}
\setsymbol{HFI:FWHM:Mars:143GHz}{7.04}
\setsymbol{HFI:FWHM:Mars:217GHz}{4.68}
\setsymbol{HFI:FWHM:Mars:353GHz}{4.43}
\setsymbol{HFI:FWHM:Mars:545GHz}{3.80}
\setsymbol{HFI:FWHM:Mars:857GHz}{3.67}

\setsymbol{HFI:beam:ellipticity:Mars:100GHz}{1.18}
\setsymbol{HFI:beam:ellipticity:Mars:143GHz}{1.03}
\setsymbol{HFI:beam:ellipticity:Mars:217GHz}{1.14}
\setsymbol{HFI:beam:ellipticity:Mars:353GHz}{1.09}
\setsymbol{HFI:beam:ellipticity:Mars:545GHz}{1.25}
\setsymbol{HFI:beam:ellipticity:Mars:857GHz}{1.03}

\setsymbol{HFI:CMB:relative:calibration:100GHz}{$\lsim 1\%$}
\setsymbol{HFI:CMB:relative:calibration:143GHz}{$\lsim 1\%$}
\setsymbol{HFI:CMB:relative:calibration:217GHz}{$\lsim 1\%$}
\setsymbol{HFI:CMB:relative:calibration:353GHz}{$\lsim 1\%$}
\setsymbol{HFI:CMB:relative:calibration:545GHz}{}
\setsymbol{HFI:CMB:relative:calibration:857GHz}{}

\setsymbol{HFI:CMB:absolute:calibration:100GHz}{$\lsim 2\%$}
\setsymbol{HFI:CMB:absolute:calibration:143GHz}{$\lsim 2\%$}
\setsymbol{HFI:CMB:absolute:calibration:217GHz}{$\lsim 2\%$}
\setsymbol{HFI:CMB:absolute:calibration:353GHz}{$\lsim 2\%$}
\setsymbol{HFI:CMB:absolute:calibration:545GHz}{}
\setsymbol{HFI:CMB:absolute:calibration:857GHz}{}

\setsymbol{HFI:FIRAS:gain:calibration:accuracy:statistical:100GHz}{}
\setsymbol{HFI:FIRAS:gain:calibration:accuracy:statistical:143GHz}{}
\setsymbol{HFI:FIRAS:gain:calibration:accuracy:statistical:217GHz}{}
\setsymbol{HFI:FIRAS:gain:calibration:accuracy:statistical:353GHz}{2.5\%}
\setsymbol{HFI:FIRAS:gain:calibration:accuracy:statistical:545GHz}{1\%}
\setsymbol{HFI:FIRAS:gain:calibration:accuracy:statistical:857GHz}{0.5\%}

\setsymbol{HFI:FIRAS:gain:calibration:accuracy:systematic:100GHz}{}
\setsymbol{HFI:FIRAS:gain:calibration:accuracy:systematic:143GHz}{}
\setsymbol{HFI:FIRAS:gain:calibration:accuracy:systematic:217GHz}{}
\setsymbol{HFI:FIRAS:gain:calibration:accuracy:systematic:353GHz}{}
\setsymbol{HFI:FIRAS:gain:calibration:accuracy:systematic:545GHz}{7\%}
\setsymbol{HFI:FIRAS:gain:calibration:accuracy:systematic:857GHz}{7\%}

\setsymbol{HFI:FIRAS:zero:point:accuracy:100GHz:units}{0.8\MJysr}
\setsymbol{HFI:FIRAS:zero:point:accuracy:143GHz:units}{}
\setsymbol{HFI:FIRAS:zero:point:accuracy:217GHz:units}{}
\setsymbol{HFI:FIRAS:zero:point:accuracy:353GHz:units}{1.4\MJysr}
\setsymbol{HFI:FIRAS:zero:point:accuracy:545GHz:units}{2.2\MJysr}
\setsymbol{HFI:FIRAS:zero:point:accuracy:857GHz:units}{1.7\MJysr}

\setsymbol{HFI:FIRAS:zero:point:accuracy:100GHz}{0.8}
\setsymbol{HFI:FIRAS:zero:point:accuracy:143GHz}{}
\setsymbol{HFI:FIRAS:zero:point:accuracy:217GHz}{}
\setsymbol{HFI:FIRAS:zero:point:accuracy:353GHz}{1.4}
\setsymbol{HFI:FIRAS:zero:point:accuracy:545GHz}{2.2}
\setsymbol{HFI:FIRAS:zero:point:accuracy:857GHz}{1.7}

\setsymbol{HFI:unit:conversion:100GHz:units}{0.2415\MJysrmK}
\setsymbol{HFI:unit:conversion:143GHz:units}{0.3694\MJysrmK}
\setsymbol{HFI:unit:conversion:217GHz:units}{0.4811\MJysrmK}
\setsymbol{HFI:unit:conversion:353GHz:units}{0.2883\MJysrmK}
\setsymbol{HFI:unit:conversion:545GHz:units}{0.05826\MJysrmK}
\setsymbol{HFI:unit:conversion:857GHz:units}{0.002238\MJysrmK}

\setsymbol{HFI:unit:conversion:100GHz}{0.2415}
\setsymbol{HFI:unit:conversion:143GHz}{0.3694}
\setsymbol{HFI:unit:conversion:217GHz}{0.4811}
\setsymbol{HFI:unit:conversion:353GHz}{0.2883}
\setsymbol{HFI:unit:conversion:545GHz}{0.05826}
\setsymbol{HFI:unit:conversion:857GHz}{0.002238}

\setsymbol{HFI:colour:correction:alpha=-2:V1.01:100GHz}{0.9893}
\setsymbol{HFI:colour:correction:alpha=-2:V1.01:143GHz}{0.9759}
\setsymbol{HFI:colour:correction:alpha=-2:V1.01:217GHz}{1.0007}
\setsymbol{HFI:colour:correction:alpha=-2:V1.01:353GHz}{1.0028}
\setsymbol{HFI:colour:correction:alpha=-2:V1.01:545GHz}{1.0019}
\setsymbol{HFI:colour:correction:alpha=-2:V1.01:857GHz}{0.9889}

\setsymbol{HFI:colour:correction:alpha=0:V1.01:100GHz}{1.0008}
\setsymbol{HFI:colour:correction:alpha=0:V1.01:143GHz}{1.0148}
\setsymbol{HFI:colour:correction:alpha=0:V1.01:217GHz}{0.9909}
\setsymbol{HFI:colour:correction:alpha=0:V1.01:353GHz}{0.9888}
\setsymbol{HFI:colour:correction:alpha=0:V1.01:545GHz}{0.9878}
\setsymbol{HFI:colour:correction:alpha=0:V1.01:857GHz}{1.0014}

\begin{document}

\title{The microwave sky after one year of \Planck\ operations}

\author{\Planck\ Collaboration, presented by A. Mennella$^*$}

\address{Dipartimento di Fisica, Universit\`a degli Studi di Milano, \\
Milano, 20133, Italy\\
$^*$E-mail: aniello.mennella@fisica.unimi.it\\
www.fisica.unimi.it}

%

\begin{abstract}
The ESA \Planck\ satellite, launched on May 14$^{\rm th}$, 2009, is the third generation space mission dedicated to the measurement of the Cosmic Microwave Background (CMB), the first light in the Universe. \Planck\ observes the full sky in nine frequency bands from 30 to 857 GHz and is designed to measure the CMB anisotropies with an unprecedented combination of sensitivity, angular resolution and control of systematic effects. In this presentation we summarise the \Planck\ instruments performance and discuss the main scientific results obtained after one year of operations in the fields of galactic and extragalactic astrophysics. 
\end{abstract}


\bodymatter

\section{Introduction}
\label{sec_introduction}

  The Cosmic Microwave Background (CMB) is constituted by relic photons that were coupled to barionic matter in the hot primordial plasma and travelled in the expanding universe when it became neutral, after $\sim$380000 years after the big bang. Today we detect it as a highly isotropic microwave background at the temperature of $\sim 2.73$\,K, with anisotropies at the level of $\Delta T/T\sim 10^{-5}$. These anisotropies trace the matter density distribution in the universe immediately before matter-radiation decoupling.

\Planck, launched on 14 May 2009, is the first European and third generation CMB space mission after COBE and WMAP; the \Planck\ instruments are designed to extract all the cosmological information encoded in the CMB temperature anisotropies with an accuracy set by cosmic variance and astrophysical confusion limits. \Planck\ will image the sky in nine frequency bands ranging from 30 to 857\,GHz, leading to a full-sky map of the CMB temperature fluctuations with signal-to-noise $>10$ and angular resolution $<10'$. In addition, all \Planck\ bands between 30 and 353\,GHz are sensitive to linear polarisation. 

\Planck\ performance is sized to map the CMB anisotropies over the entire angular range dominated by primordial fluctuations. This will lead to accurate estimates of cosmological parameters that describe the geometry, dynamics, and matter-energy content of the universe. The \Planck\ polarisation measurements are expected to deliver complementary information on cosmological parameters and to provide a unique probe of the thermal history of the universe in the early phase of structure formation. \Planck\ will also test the inflationary paradigm with unprecedented sensitivity through studies of non-Gaussianity and of B-mode polarisation as a signature of primordial gravitational waves. 

The wide frequency range of \Planck\ is required primarily to ensure accurate discrimination of foreground emissions from the cosmological signal. However, the nine maps also represent a rich data set for galactic and extragalactic astrophysics. In this paper we present an overview of the status of the \Planck\ mission and of the preliminary results in the field of galactic and extragalactic astrophysics obtained after one year of operations.

\section{The \Planck\ satellite and its operations}
\label{sec_operations}

%
%

\begin{figure}[h!]
  \begin{center}
    \includegraphics[width=5cm]{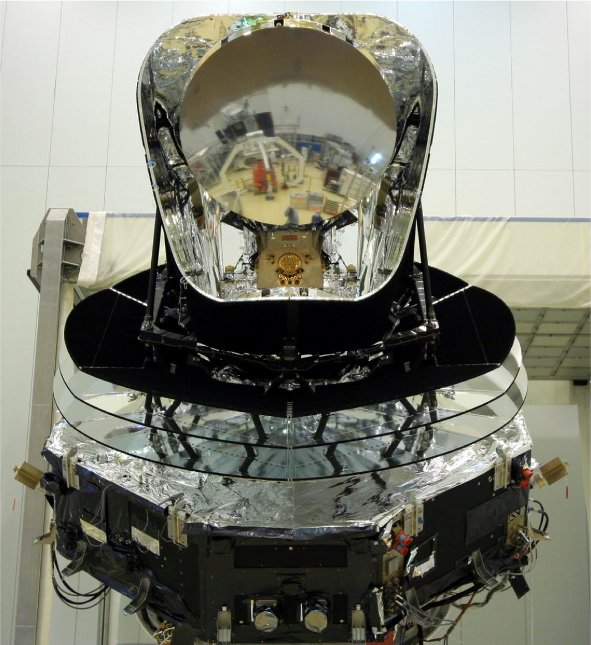}\includegraphics[width=6.5cm]{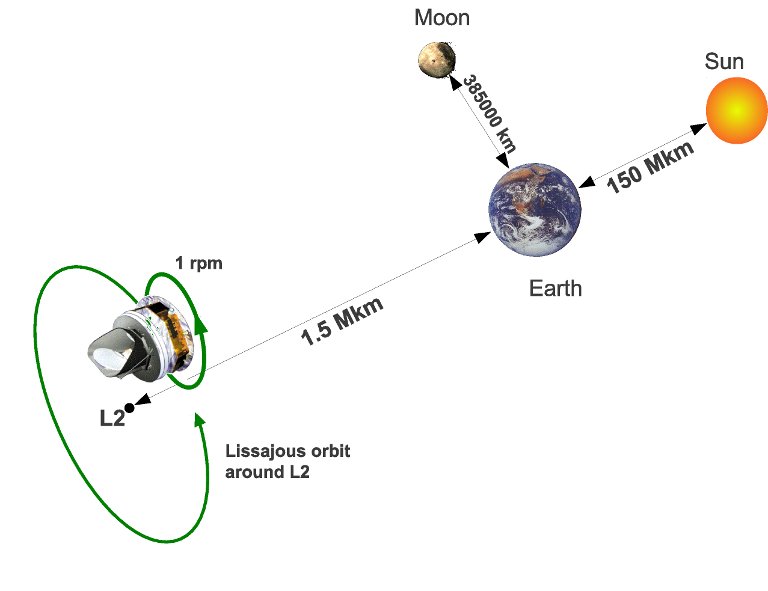}
  \end{center}
  \caption{Left panel: the \Planck\ satellite a few days before launch. Right panel: \Planck\ orbits around the second Lagrangian point, L2, and scans the sky in near-great circles by spinning at 1 r.p.m. The satellite is repointed approximately hourly by 2$'$ so that the solar aspect angle is kept constant and the full sky is observed in about six months.}
  \label{fig_planck_satellite}
\end{figure}

\Planck\  (see left panel of  Fig.~\ref{fig_planck_satellite}) is a spinner constituted of two modules\cite{tauber2010a}: (i) a payload module containing telescope, instruments, a baffle that provides straylight rejection and radiative cooling, and three conical ``V-groove'' radiators that thermally decouple the warm and cold satellite modules; (ii) a service module containing the warm satellite and instrument electronics, the solar cells, the cryo-coolers, the main on-board computer, the telecommand receivers and telemetry transmitters, and the attitude control system with its sensors and actuators.

The telescope\cite{tauber2010b} is a dual-reflector off-axis aplanatic Gregorian telescope with 1.5\,m primary projected aperture, pointing at 85$^\circ$ with respect to the spin axis. It focusses the sky radiation on the secondary mirror focal plane, hosting the feed horn antennas of the two instruments: the Low Frequency Instrument (LFI) and the High Frequency Instrument (HFI).

\Planck\ was launched together with the ESA's Herschel observatory on 14 May 2009 (13:12 UT) from the Centre Spatial Guyanais in Kourou (French Guyana) on an Ariane 5 ECA rocket\cite{planck2011-1.1}, and was placed in its final Lissajous orbit around the second Lagrangian point of the Earth-Sun system (``L2'') after three large manoeuvres. The right panel in Fig.~\ref{fig_planck_satellite} sketches the orbit and provides the basic details of the scanning strategy.

\Planck\ started cooling down radiatively shortly after launch. Approximately two months were necessary to complete the cooldown and reach the nominal temperatures of the \Planck\ cryogenic stages\cite{planck2011-1.3}: (i) 50\,K, reached by passive cooling, for the telescope, baffle and upper V-groove, (ii) 20\,K provided by the Hydrogen \Planck\ Sorption Cooler, for the LFI focal plane and HFI pre-cooling, (iii) 4.5\,K provided by a $^4$He-JT Stirling cooler, for the HFI focal plane feeds and LFI reference loads and (iv) 1.6\,K and 0.1\,K, provided by a $^3$He-$^4$He open cycle dilution cooler, for the HFI filters and bolometers, respectively.

During cooldown, several activities were carried out to verify instrument functionality, tune the main parameters and check that scientific performance was comparable with ground measurements\cite{planck2011-1.3,planck2011-1.4}. At the end of this phase the two instrument were fully tuned and ready for routine operations. On 13 August 2009 \Planck\ entered the so-called ``First Light Survey'' (FLS), a two-week period during which \Planck\ operated as if it were in its routine phase. The conclusion was that the \Planck\ payload required no further tuning of its instruments, so that the FLS was accepted as a valid part of the first \Planck\ survey.

The instruments have continued working smoothly and their parameters have not been changed ever since. The only notable event has been the planned switch-over from the nominal to the redundant Sorption Cooler which happened during August 2010. Nominal operations will continue until the end of the dilution cooler refrigerant, which is foreseen by the end of January 2012. At that time the HFI will become not operative and the LFI will continue scanning the sky for up to another year, depending on the residual duration of the Sorption Cooler lifetime.

\section{Instruments and scientific performance}
\label{sec_instruments_performance}

\begin{figure}[h!]
    \begin{center}
	\includegraphics[width=11cm]{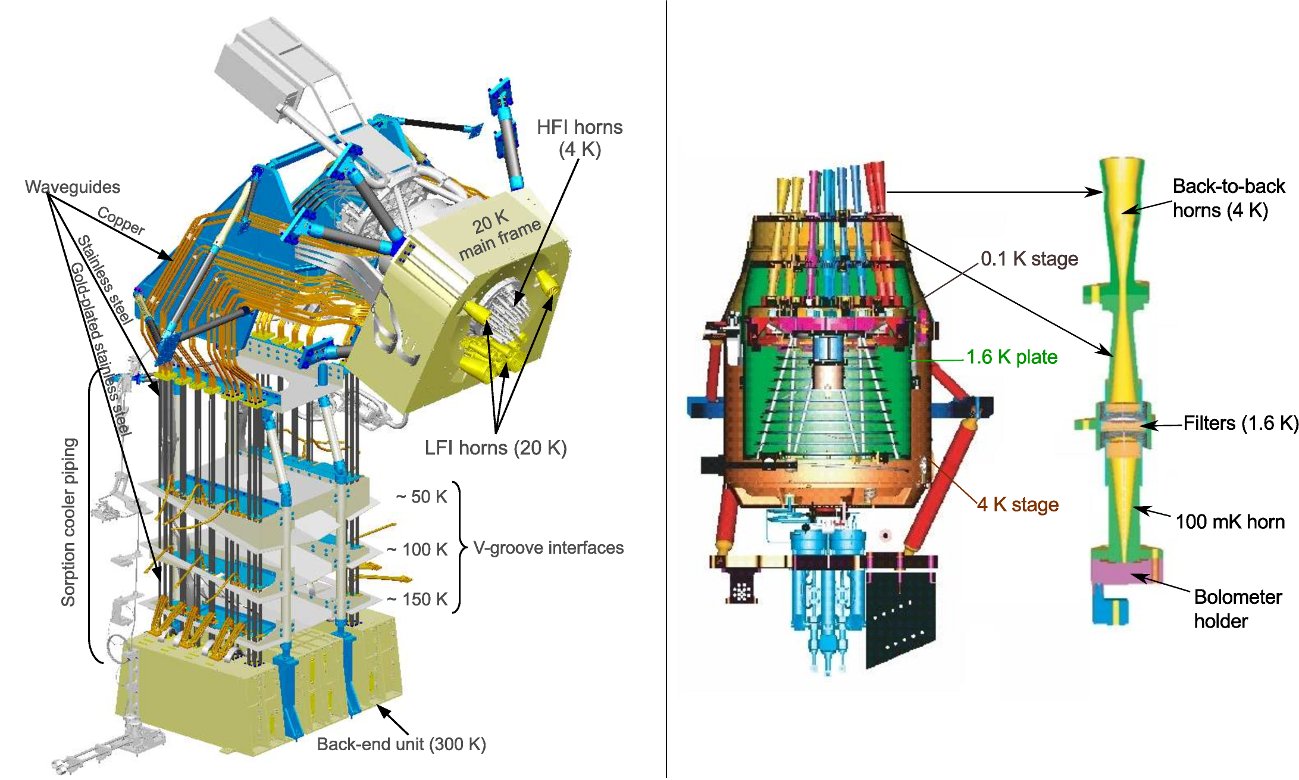}
    \end{center}
    \caption{Left panel: 3-D view of the LFI instrument. Right panel: the HFI focal plane and a schematic of the bolometric optical chain.}
    \label{fig_lfi_hfi_instrument}
  \end{figure}

  The LFI\cite{bersanelli2010} (left panel of Fig.~\ref{fig_lfi_hfi_instrument}) is an array of 11 microwave coherent pseudo-correlation differential receivers in Ka, Q and V bands. They are based on low noise amplifiers using indium-phosphide high electron mobility transistors, operated at a temperature of $\sim$20 K. The HFI\cite{lamarre2010} (right panel of Fig.~\ref{fig_lfi_hfi_instrument}) is designed around 52 bolometers operated at 0.1\,K and fed by corrugated feedhorns and bandpass filters within a back-to-back conical horn optical waveguide. Twenty of the bolometers are sensitive to total power, and the remaining 32 units are arranged in pairs of orthogonally-oriented polarisation-sensitive bolometers (PSBs). 

The in-flight measured main performance parameters of the \Planck\ instruments are summarised in Table~\ref{tab_planck_performance}.

\begin{table}
    \tbl{\Planck\ performance parameters determined from flight data.}
{\begin{tabular}{l c c c c c c c}
  \hline
  \multicolumn{5}{c}{\mbox{ }} & \multicolumn{2}{c}{W{\sc hite-noise}} & \\
  \multicolumn{3}{c}{\mbox{ }} & \multicolumn{2}{c}{M{\sc ean} B{\sc eam}} & \multicolumn{2}{c}{S{\sc ensitivity}} & {C{\sc alibr.}}\\
    & & $\nu_{\rm center}$ & \multicolumn{2}{c}{\hrulefill} &\multicolumn{2}{c}{\hrulefill} &{U{\sc nc.}} \\
   C{\sc hannel} &$N_{\rm detectors}$ &[GHz]&FWHM&Ellipticity&[\muKRJs]&[\muKCMBs]&[\%]\\
   \hline
  \phantom{*}30\,GHz&\phantom{*}4&\phantom{**}\phantom{.}\getsymbol{LFI:center:frequency:30GHz}&\getsymbol{LFI:FWHM:30GHz}& 1.38&\getsymbol{LFI:white:noise:sensitivity:30GHz}&146.8&1\\
  \phantom{*}44\,GHz&\phantom{*}6&\phantom{**}\phantom{.}\getsymbol{LFI:center:frequency:44GHz}&\getsymbol{LFI:FWHM:44GHz}& 1.26&\getsymbol{LFI:white:noise:sensitivity:44GHz}&173.1&1\\
  \phantom{*}70\,GHz&12&\phantom{**}\phantom{.}\getsymbol{LFI:center:frequency:70GHz}&\getsymbol{LFI:FWHM:70GHz}& 1.27&\getsymbol{LFI:white:noise:sensitivity:70GHz}&152.6&1\\
  100\,GHz&\phantom{*}8&\getsymbol{HFI:center:frequency:100GHz}& \phantom{*}\getsymbol{HFI:FWHM:Mars:100GHz}&\getsymbol{HFI:beam:ellipticity:Mars:100GHz}&\phantom{*}17.3&\phantom{*}22.6&2\\
  143\,GHz&11&\getsymbol{HFI:center:frequency:143GHz}& \phantom{*}\getsymbol{HFI:FWHM:Mars:143GHz}&\getsymbol{HFI:beam:ellipticity:Mars:143GHz}&\phantom{*}\phantom{*}8.6&\phantom{*}14.5&2\\
  217\,GHz&12&\getsymbol{HFI:center:frequency:217GHz}& \phantom{*}\getsymbol{HFI:FWHM:Mars:217GHz}&\getsymbol{HFI:beam:ellipticity:Mars:217GHz}&\phantom{**}6.8&\phantom{*}20.6&2\\
  353\,GHz&12&\getsymbol{HFI:center:frequency:353GHz}& \phantom{*}\getsymbol{HFI:FWHM:Mars:353GHz}&\getsymbol{HFI:beam:ellipticity:Mars:353GHz}&\phantom{**}5.5&\phantom{*}77.3&2\\
  545\,GHz&\phantom{*}3&\getsymbol{HFI:center:frequency:545GHz}& \phantom{*}\getsymbol{HFI:FWHM:Mars:545GHz}&\getsymbol{HFI:beam:ellipticity:Mars:545GHz}&\phantom{**}4.9&\dots&7\\
  857\,GHz&\phantom{*}3&\getsymbol{HFI:center:frequency:857GHz}& \phantom{*}\getsymbol{HFI:FWHM:Mars:857GHz}&\getsymbol{HFI:beam:ellipticity:Mars:857GHz}&\phantom{**}2.1&\dots&7\\
  \hline
\end{tabular}}\label{tab_planck_performance}
\end{table}

\subsection{The Low Frequency Instrument}
\label{sec_lfi}

  The instrument consists of a $\sim 20$\,K focal plane unit hosting the corrugated feed horns, orthomode transducers (OMTs), and receiver front-end modules (FEMs). A set of 44~composite waveguides interfaced with the three V-groove radiators \cite{planck2011-1.3} connects the front-end modules to the warm ($\sim 300$\,K) back-end unit (BEU), which contains further radio frequency amplification, detector diodes, and electronics for data acquisition and bias supply. 
    
  In each receiver the feed horn is connected to an OMT, which splits the incoming radiation into two perpendicular linear polarisation components that propagate through two independent pseudo-correlation differential radiometers. In each radiometer, the sky signal is continuously compared with a stable 4.5\,K reference load mounted on the external shield of the HFI 4\,K box. After being summed by a first hybrid coupler, the two signals are amplified by $\sim 30$\,\hbox{dB}. A phase shift alternating at 4096\,Hz between 0\deg\ and 180\deg\ is applied in one of the two amplification chains.  A second hybrid coupler separates back the sky and reference load components, which are further amplified, detected and digitised in the warm \hbox{BEU}. After the digital conversion, data are downsampled, requantised and assembled into telemetry packets.

  The LFI differential design is similar to the WMAP scheme\cite{jarosik2003} but it uses an internal blackbody load as a stable reference. The receiver is balanced via a gain modulator factor that provides good suppression of systematic effects and 1/$f$ noise knee frequencies of $\sim 50$\,mHz\cite{planck2011-1.4}.

  Data are continuously calibrated using the dipole modulation in the \hbox{CMB}. In particular our calibration signal is the sum of the solar dipole and the orbital dipole, which is the contribution from \Planck's orbital velocity around the Sun. This model allows us to reconstruct the mean value of the calibration constant with an accuracy better than 1\%. Variations in the radiometer gain on timescales of the order of few days are traced with an accuracy better than 0.1\%\cite{planck2011-1.6}.

\subsection{The High Frequency Instrument}
\label{sec_hfi}

  The HFI is a bolometric array designed to reach photon-noise limited sensitivity. Its architecture is based on independent optical chains collecting the light from the telescope and feeding it to bolometric detectors. Sixteen channels (at 100, 143, 217 and 353\,GHz) feed pairs of polarisation-sensitive bolometers (PSBs), while the remaining 20 channels (at 143, 217, 353, 545 and 857\, GHz) feed spider-web bolometers (SWBs), which are not sensitive to polarisation.

  A typical optical chain is shown in the right panel of Fig.~\ref{fig_lfi_hfi_instrument}. At 4\,K, the back-to-back horns provide initial geometrical and spectral selection of the radiation, and a first set of filters blocks the highest frequency and most energetic part of the background. In order to ensure proper positioning and cooling, these elements are attached onto three thermal stages (at 4.5\,K, 1.4\,K and 0.1\,K) in a nested arrangement. The six HFI spectral bands cover all frequencies from 84\,GHz up to 1\,THz via adjacent bands with close to 33\% relative bandwidth.
  
  Bolometers consist of (i) an absorber that transforms the incoming radiation into heat; (ii) a thermometer that is thermally linked to the absorber and measures the temperature changes; and (iii) a weak thermal link to a thermal sink, to which the bolometer is attached. In the SWBs\cite{bock1995,mauskopf1997}, the absorbers consist of metallic grids deposited on a Si3 N4 substrate in the shape of a spider web. The absorber of PSBs is a rectangular grid with metallization in one direction\cite{jones2003}. Electrical fields parallel to this direction develop currents and then deposit some power in the grid, while perpendicular electrical fields propagate through the grid without significant interaction. A second PSB perpendicular to the first one absorbs the other polarisation. Such a PSB pair measures two polarisations of radiation collected by the same horns and filtered by the same devices.

  The readout electronics consist of 72 channels that perform impedance measurements of 52 bolometers, two blind bolometers and 16 accurate low temperature thermometers. It is split into three boxes: the 50\,K JFET box, the 300\,K pre-amplifier unit (PAU), which provides an amplification of the low level voltages by a factor of 1000, and the 300\,K readout electronic unit (REU), which provides a further variable gain amplification and contains all the interfaces between the analogue and the digital electronics. Digital signal are then processed in the data processing unit (DPU) that downsample and compress data before sending to telemetry.
  
  The primary absolute calibration of HFI is based on extended sources. At low frequencies (353\,GHz and below), the orbital dipole and the Solar dipole provide good absolute calibrators. At high frequencies (545\,GHz and above), Galactic emission is used. 
  
  Bolometer data are affected by by glitches resulting from the interaction of cosmic rays with the bolometers. Because the glitch rate is of typically one per second and many of them are characterised by long decay times (from some tens of ms up to 2\,s), simple flagging would lead to an unacceptably large loss of data. Therefore only the initial part of the glitch is flagged and discarded, while the tail is fit with a model which is subtracted from the TOI. The temporal redundancy of the sky observations is used to separate the glitch signal from the sky signal.

\section{The \Planck\ microwave sky}
\label{sec_first_light}

\begin{figure}[h!]
  \begin{center}
      \includegraphics[width=11cm]{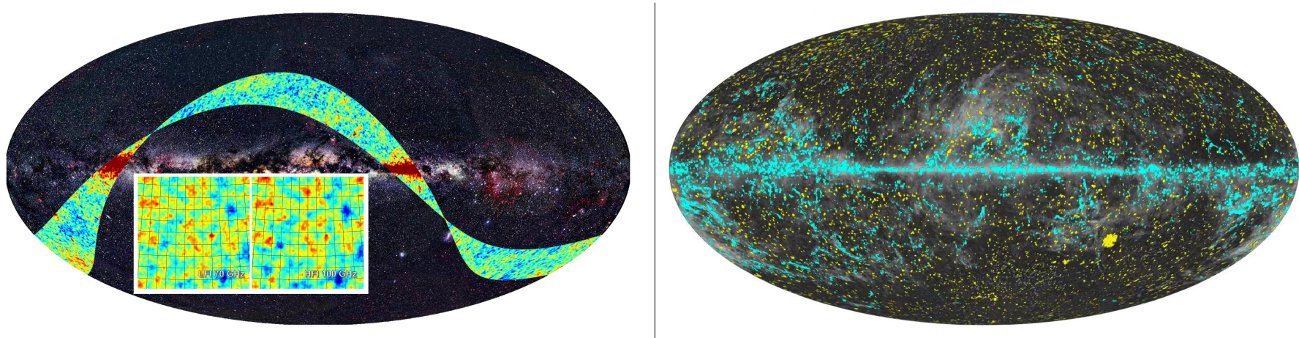}
  \end{center}
  \caption{Left: the \Planck\ First Light Survey. The 14 days survey (data from LFI 70 GHz channel) is superimposed to an optical image of the whole sky. Two insets show details of a region dominated by the CMB anisotropies in the 70 and 100\,GHz channels, that are less sensitive to foregrounds contamination. Right: the set of \Planck\ compact sources (galactic -- cyan, extragalactic -- yellow) detected during the first year of operations.}
  \label{fig_first_light_ercsc}
\end{figure}

  The First Light Survey started on 13 August 2009 and was completed on 27 August, yielding maps of a strip of the sky, one for each of \Planck's nine frequencies. Each map is a ring about 15 degrees wide, stretching across the full sky. The coloured strip in the left panel of Fig.~\ref{fig_first_light_ercsc} represents data from the 70\,GHz channel (the most insensitive to foregrounds contamination) superimposed to an optical image of the whole sky. The two insets show a relatively foreground-free sky patch for the LFI-70\,GHz (left) and HFI-100\,GHz (right) \Planck\ channels. The CMB anisotropy structure is clearly visible at high galactic latitudes and in the two insets, showing consistency in the data from the two frequency channels.

On 11 January 2011 the \Planck\ Collaboration released its first set of scientific data to the public: the Early Release Compact Source Catalogue (ERCSC), a list of more than 15000 unresolved and compact sources extracted from the first complete all-sky survey (see right panel of Fig.~\ref{fig_first_light_ercsc}). The ERCSC\cite{planck2011-1.10} consists of nine lists of sources, extracted independently from each of \Planck's frequency bands, and two lists of sources extracted using multi-band criteria: (i) ``Cold Cores,'' cold and dense locations in the Insterstellar Medium of the Milky Way, and (ii) clusters of galaxies, selected using the spectral signature left on the Cosmic Microwave Background by the Sunyaev-Zeldovich (SZ) effect.

\begin{figure}[h!]
  \begin{center}
      \includegraphics[width=11.5cm]{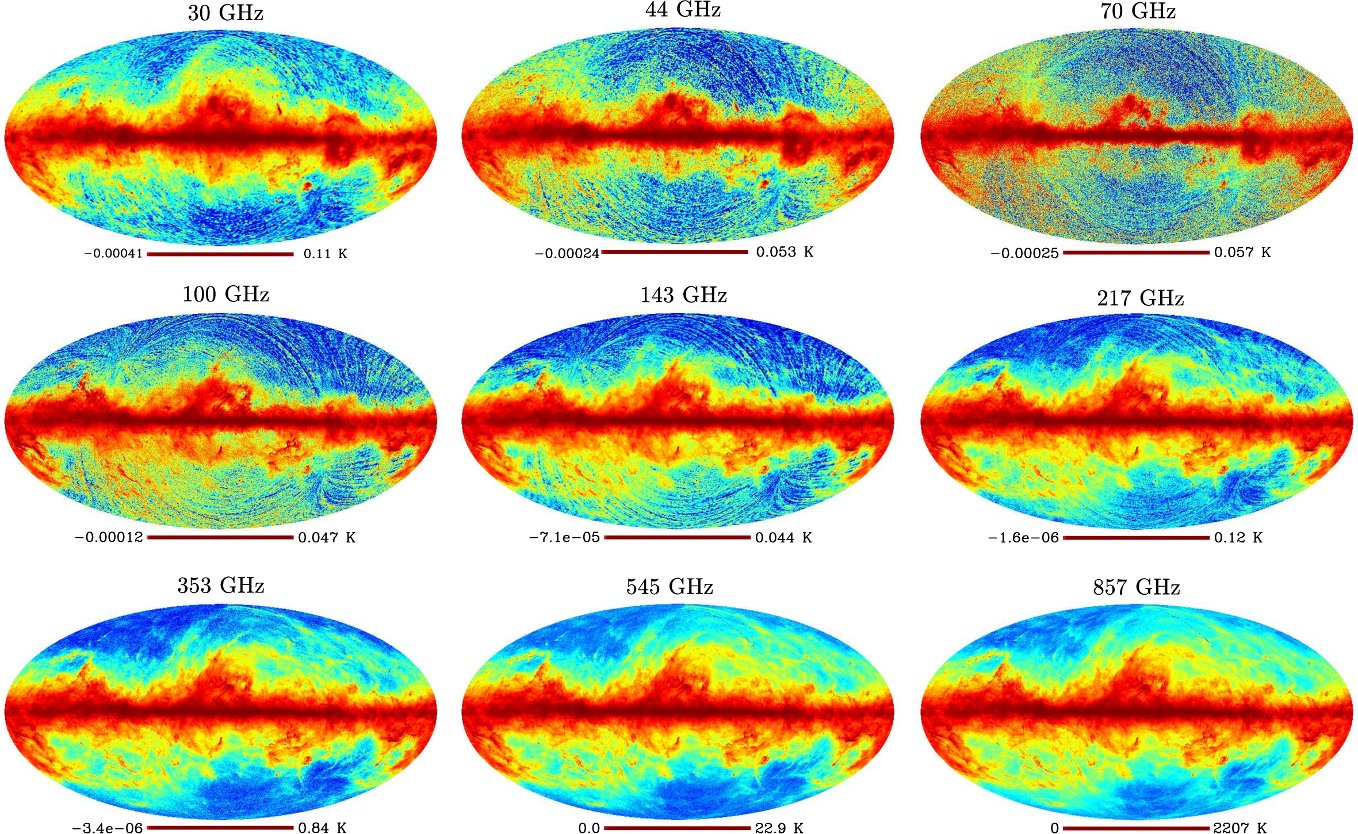}
  \end{center}
  \caption{The nine \Planck\ foreground full-sky maps.}
  \label{fig_frequency_maps}
\end{figure}

  
In Figure~\ref{fig_frequency_maps} we show the full set of maps of the astrophysical foregrounds at the nine \Planck\ frequencies, where the CMB component has been removed as described in Refs.~[\refcite{planck2011-1.6,planck2011-1.7}]. This set of maps represents the first full-sky view of the microwave sky ever observed by a single experiment in such a wide frequency span.  

These nine maps represent a key element in the data processing pipeline that extracts the CMB anisotropies separating them from the foreground signals and also provide a wealth of new information shedding light on many open questions in galactic and extragalactic astrophysics. A set of 18 papers have been submitted by the \Planck\ team together with the release of the ERCSC discussing preliminary scientific results obtained from the catalogue and from the maps which were used as input for the production of the ERCSC. 

Although covering the details of such studies is outside the scope of this paper (see the complete list of references provided in [\refcite{planck2011-1.1}]) we outline here the main areas of scientific interest:

\begin{itemize}
  \item detection of galaxy clusters via the Sunyaev-Zeldovich (SZ) effect and correlation with X-ray follow-up observations (e.g. with the XMM-Newton X-ray observatory);
  \item statistics of the spectral distribution of radio sources in the microwave region;
  \item detection and statistics of ``Cold Cores'', i.e. regions of cold dust gravitationally collapsing in star-forming regions;
  \item study of the physical properties of the interstellar medium (``dark gas'', i.e. gas not spatially correlated with known tracers of neutral and molecular gas, molecular clouds, anomalous emission from interstellar dust which can be interpreted as arising from small spinning grains).
\end{itemize}

\section{Next steps}
\label{sec_next_steps}

  At the time of writing \Planck\ has completed the fourth sky survey and it is entering the fifth. The two instruments have so far maintained their initial performance and are expected to continue observing the sky until January 2012. As expected, at that time the HFI open-cycle dilution cooler will run out of the necessary Helium refrigerant and the temperature of the bolometric detectors will rise to $\sim$4.5\,K. At this temperature the bolometers will become essentially blind to the CMB photons and the HFI will end its operations.

An extended operation period is currently foreseen for the LFI, with an approved duration up to one additional year, depending on the residual lifetime of the \Planck\ Sorption Cooler.

The next release of \Planck\ products will take place in January 2013, and will cover data acquired in the period up to 27 November 2010. At this time the first \Planck\ cosmological results will be released together with the necessary data to support them. A third release of products is foreseen after January 2014, to cover the data acquired beyond November 2010 and until the end of \Planck\ operations.

\section{Conclusions}
\label{sec_conclusions}

  \Planck, the third space mission dedicated to the measurements of the CMB anisotropies, has been successfully scanning the sky for more than two years and has started its fifth sky survey as we write. Its focal plane instruments are the most sensitive microwave detectors ever flown and have confirmed and maintained their predicted performance.

The first public \Planck\ data release, on 11 January 2011, has provided the scientific community with a preliminary full-sky catalogue of compact, unresolved sources in a largely unexplored frequency range. These early Planck results have already started shedding new light on several astrophysical issues like detection of new galaxy clusters, the physics of stellar formation in cold gas regions, the nature of ``dark gas'', etc. The first release of cosmological data and results is foreseen by January 2013, and a second one, based on the complete \Planck\ dataset, will happen in early 2014.

After 19 years \Planck\ is approaching its end of operations but the data analysis is currently at full speed to provide the scientific community with the highest quality CMB data ever released. The best still has to come.
  
\section*{Acknowledgements}

  \Planck\ is operated by ESA via its Mission Operations Centre located at ESOC (Darmstadt, Germany). ESA also coordinates scientific operations via the \Planck\ Science Office located at ESAC (Madrid, Spain). Two Consortia, comprising around 50 scientific institutes within Europe, the USA, and Canada, and funded by agencies from the participating countries, developed the scientific instruments LFI and HFI, and continue to operate them via Instrument Operations Teams located in Trieste (Italy) and Orsay (France). The Consortia are led by the Principal Investigators: J.-L. Puget in France for HFI (funded principally by CNES and CNRS/INSU-IN2P3) and N. Mandolesi in Italy for LFI (funded principally via ASI). NASA's US \Planck\ Project, based at JPL and involving scientists at many US institutions, contributes significantly to the efforts of these two Consortia. In Finland, the \Planck\ LFI 70 GHz work was supported by the Finnish Funding Agency for Technology and Innovation (Tekes). This work was also supported by the Academy of Finland, CSC, and DEISA (EU).

\bibliographystyle{pippo}
\bibliography{bibliography}

\end{document}